# What Risk Factors to Cause Long COVID and Its Impact on Patient Survival Outcomes when Combined with the Effect from Organ Transplantation in the Acute COVID

Ahmed N Nyandemoh[1] MPH, Jerrod Anzalone[2] MPH, Hongyin Dai[1] Ph.D, Roslyn Mannon[3] MD, and Jianghu (James) Dong[1, 3, *] Ph.D

1  Department of Biostatistics, College of Public Health, University of Nebraska Medical Center
2  Department of Neurological Sciences, University of Nebraska Medical Center
3  Division of Nephrology, Department of Internal Medicine, University of Nebraska Medical Center

*Corresponding Author:

Jianghu (James) Dong, Ph.D.

Department of Biostatistics, College of Public Health

University of Nebraska Medical Center

984375 Nebraska Medical Center, Omaha, NE 68198-4375

p: 402.559.1976 | f: 402.559.4961

Jianghu.dong@unmc.edu

**Abstract:**

**Background:** Coronavirus disease 2019 (COVID-19) in solid organ transplant (SOT) patients is associated with more severe outcomes than non-immunosuppressed hosts. However, exactly which risk factors cause Long COVID in acute COVID cases remains unknown. More importantly, the impact of Long COVID on patient survival remains understudied, especially when examined alongside the effect of SOT.

**Methods:** All patients have been identified with acute COVID in the National COVID Collaborative Cohort registry from July 1, 2020, to June 30, 2022. We compared patient demographics in Long COVID vs. those without Long COVID based on descriptive statistics. Multivariable logistics regressions were used to determine the factors related to the likelihood of developing Long COVID from a case of acute COVID. Multi-variables Cox regression was used to determine the time-to-event outcome of patient survival with Long COVID.

**Results:** This study reviewed data from a cohort of 6,416,500 acute COVID patients. Of that group, 31,744 (0.5%) patients developed Long COVID from ICD diagnosis. The mean (q1, q3) age was 39 (22, 57) years old, and 55% of patients were female. From this cohort, a total of 31,744 (1%) developed Long COVID and 43,565 (1%) had SOT, with a total of 698 SOT patients identified with Long COVID. Mean age of those with Long COVID was 52 (39, 64) years old and 64% of patients were female. Most of the SOT patients were kidney transplant recipients. From the Cox regression analysis of patient survival, there were many significant factors related to patient survival (death), with elderly SOT patients having a much higher hazard ratio of 27.8 (26.3, 29.4).

**Conclusion:** This study has identified the important risk factors that are more likely to cause Long COVID in an acute COVID cohort. We investigated hazard ratios of patient survival based on multivariable Cox models, which found that Long COVID had a more direct impact on survival in elderly patients and those with SOT.

**Introduction**

Long COVID, also known as post-COVID conditions (PCC) or post-acute sequelae of COVID-19 (PASC), is the common term used to describe signs and symptoms that last for longer than four weeks after cases of acute COVID-19. Long COVID is a novel syndrome that is broadly defined by the persistence of physical, psychological, or cognitive symptoms following a probable or confirmed SARS-CoV-2 infection. Available evidence suggests that Long COVID is a substantial public health problem with severe consequences for affected individuals and society at large. Patients commonly report being emotionally affected by health problems related to Long COVID. In the United States, patients have reported mild to severe financial impacts related to COVID-19 in multiple studies [1,2,3]. This concern is underlined by reports that Long COVID patients experience increased disability related to breathlessness and decreased quality of life [4]. Understanding the needs of these patients will allow for the development of healthcare, rehabilitation, and other resources needed to support their recovery [5,6]

The COVID-19 pandemic's impact on solid organ transplantation (SOT) recipients has been reflective. The SOT patients with COVID-19 appear to be at higher risk of severe outcomes based on their chronically immunosuppressed state and underlying medical comorbidities [7,8,9]. Most SOT patients have one or several associated risk factors for severe COVID-19 or death, such as hypertension, cardiovascular disease, and chronic kidney disease. Hence, some studies reported that SOT patients had a higher COVID-19-related mortality rate than non-transplant patients[3,10,11]. Due to the novelty of COVID-19, there is a lack of data regarding many aspects, including its natural course within immunocompromised hosts, the utility of current treatment regimens employed in non-transplant individuals for SOT patients, and the effect of immunosuppression on the course of the disease[12,13]. Even though comorbidities, such as obesity, have been reported as risk factors for severe disease [14], an immunocompromised state has not yet been proven to have a worse clinical outcome than SARS-CoV-2 infection. However, SOT patients with Long COVID are particularly unknown due to a high prevalence of underlying chronic kidney disease, diabetes mellitus, hypertension, and calcineurin inhibitor (CNI) use.

Therefore, this retrospective cohort study was conducted using data available from the National COVID Cohort Collaborative (N3C) program to address these research questions. This study aimed to synthesize existing evidence on the demographics, comorbidities, and SOT aspects of the incidence of Long COVID and patient survival in acute COVID-19 patients, as well as its prevalence. Information from this study could be used to impact transplant patients with long COVID and acute COVID by identifying gaps in demographics and clinical characteristics within the study period. We will identify the impact and potential risk factors for the Long COVID event and SOT effect, and address patient survival in those with Long COVID.

**Patients and Methods**

**Data resource**

The N3C Data Enclave is a secure platform through which the clinical data provided by contributing members is stored in a centralized, secure format. The N3C platform is a multi-center, federated research network in the United States that provides real-time access to de-identified healthcare record data of more than 16 million patients from participating healthcare organizations. The data itself can only be accessed through a secure cloud portal hosted by NCATS and cannot be downloaded or removed.

The retrospective cohort study was conducted using data from the N3C platform, as it is a health record repository containing the largest, most representative U.S. cohort of COVID-19 cases and controls to date. In the real-time search, the source data were collected as part of routine healthcare and is historical in nature. The data was extracted and analyzed using Python, SQL, and R in the N3C platform. The study evaluated pre-existing medical conditions, and the research included all patients with acute COVID.

The acute COVID patients were determined *a priori* and included patient demographics, such as age, race, sex, type of organ transplant (multiple transplantations, kidney transplantation, liver transplantation, heart transplantation, lung transplantation, and other/unknown transplantations), and comorbidities (hypertension, diabetes, asthma, malignant cancer, coronary artery disease, congestive heart failure, and peripheral vascular disease).

**Statistical Analyses**

We used descriptive statistics, such as mean (SD) or median (q1 and q3), as appropriate to report the baseline demographic and clinical characteristics by subgroups, where all patients were stratified by whether they had developed Long COVID or not. Counts and percentages for demographic and clinical characteristics were used in the study population to describe categorical variables. The odds ratio of whether receiving a Long COVID event was determined based on multivariable logistic regressions. Finally, the following co-variables, including age (i.e., < 19, 19–34, 35–49, 50–64, and 65+ years), sex subgroups, and transplant type (multiple, kidney, heart, lung, liver), were used in Cox regression models for patient survival outcome. The results of the Cox regression analyses were displayed using hazard ratios (HRs) and 95% confidence intervals (CIs). P-values lower than 0.05 were considered statistically significant in the statistical models.

**Results**

**Patient Demographic**

The time period in this study was chosen from July 1st, 2020, to June 30th, 2022. A total of 6,416,500 patients were found to have acute COVID, with a subset of 31,744 (0.5%) of those patients developing Long COVID (from ICD diagnosis) and 43,565 (1%) having had a SOT. The mean (q1, q3) age was 39 (22, 57) years old, and 55% of patients were female. A total of 698 individuals were identified to be transplant patients with Long COVID. The demographic comparison of these patients is shown in Table 1. Hypertension before COVID indicators, diabetes before COVID indicators, congestive heart failure before and post COVID indicators, coronary artery disease before and post COVID indicators, and peripheral vascular disease before and post COVID indicators were all common comorbidities; however, these were significantly more common in those who developed Long COVID.

Those who had acute COVID were significantly more likely to be kidney transplant recipients relative to other organ transplant types (58% of acute COVID versus 51% of Long COVID SOT patients had a kidney transplant, P value < 0.001) because most SOT patients are kidney transplant recipients. It is interesting to find that liver transplant recipients had 15% of acute COVID versus 9.8% of Long COVID (P value < 0.001). Conversely, lung (7.0% vs. 13.0%, P value< 0.001) and heart (10% vs. 13.0%) transplant recipients made up a smaller proportion of those patients with acute COVID than those with Long COVID. The categorical transplant variables include kidney transplant (61%) as the highest of the transplants, followed by liver transplant (19%), lung transplant (18%), and heart transplant (16%), as shown in Figure 1.

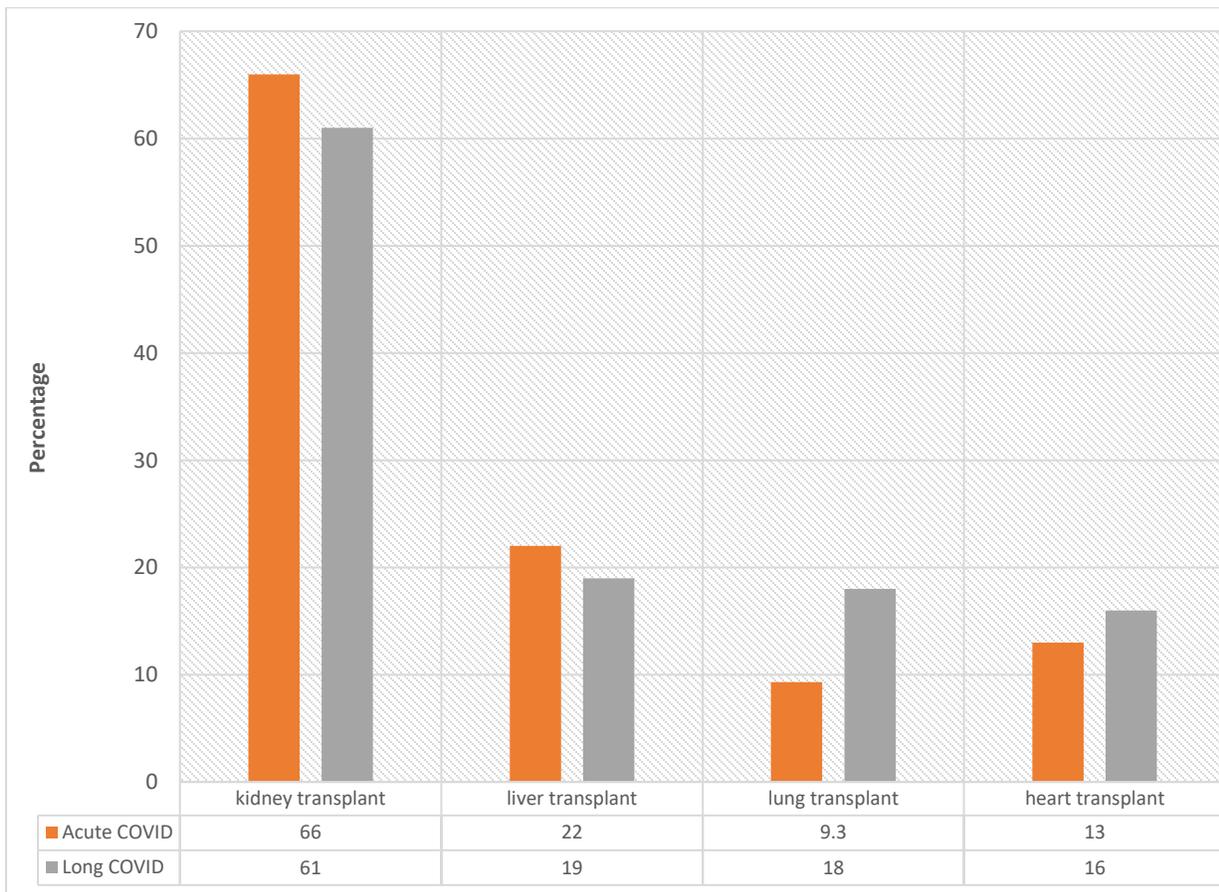

*Figure 1*: Among SOT transplants with acute COVID and long COVID, most patients are kidney transplant recipients. But the number of lung transplant patients with Long COVID was double the number of lung transplant patients with acute COVID.

All patients for comorbidities reported whether there was a statistical difference based on P-values. Hypertension, diabetes, coronary artery disease, peripheral vascular disease, and congestive heart failure were common comorbidities in the SOT cohort, but were significantly more common in those who had Long COVID and acute COVID.

Further analysis in more recent cohorts can be done through some sensitivity analysis by the inclusion of acute COVID without Long COVID individuals matched by age, sex, and race.

**Results from Statistical Models**

In the acute COVID cohort, results of the logistics regression analysis of long-term COVID is shown in Table 2, with a higher odds ratio (OR) of Long COVID present amongst patients older than 65 years (OR:1.73, 95% CI: 1.66–1.81), with hypertension (OR: 1.36, 95% CI: 1.32–1.40), with diabetes (OR: 1.09, 95% CI: 1.05–1.14), with asthma (OR:1.85, 95% CI: 1.79–1.91), with malignant cancer (OR: 1.12, 95% CI: 1.07–1.16), with congestive heart failure (OR: 1.27, 95% CI: 1.22–1.32), with peripheral vascular disease (OR: 1.02, 95% CI: 0.96–1.07), and with dementia (OR: 1.18, 95% CI: 1.08–1.29). The results of patient survival from the Cox regression models are shown in Table 3. Females were less likely to die, as seen by the hazard ratio (HR = 0.69; 95% CI: 0.68–0.70) while adjusting for Male patients. The HR of age categories ranged from 2.74 in patients aged < 19 years (95% CI: 2.57–2.92) to 27.8 in elderly patients over 65 years old (95% CI: 26.3–29.4). The detrimental effect of Long COVID in elderly SOT patients can be observed as the hazard ratio was so high at 27.8 (26.3, 29.4).

**Conclusion**

This study utilizes the largest, most representative U.S. cohort of COVID-19 cases and controls to date, evaluating the risk of Long COVID and patient outcomes from patients with acute COVID. In this cohort of patients followed over a two-year period, 1% developed Long COVID and 2% had SOT. Advanced age and having had a transplant were associated with an increased impact on patient survival.

This study identified risk factors for developing Long COVID from an acute COVID and SOT patient cohort. For the SOT cohort, we also investigated hazard ratios of patient survival based on multivariable Cox models. From this, we find that elderly SOT patients were much more likely to die from Long COVID.

Table 1: Patient demographics and clinical characteristics among Long COVID patients versus acute COVID patients without Long COVID

| Characteristic variables | Overall | Acute COVID without Long COVID | Long COVID | P value |
|---|---|---|---|---|
| Gender | | | | <0.001 |
| Male | 2872550 (45%) | 2861210 (45%) | 11340 (36%) | |
| Female | 3543950 (55%) | 3523546 (55%) | 20404 (64%) | |
| Mean (Q₁, Q₃) of age | 39 (22, 57) | 39 (22, 57) | 52 (39, 64) | <0.001 |
| <19 | 1269106 (20%) | 1267444 (20%) | 1662 (5%) | |
| 19-34 | 1503840 (23%) | 1499755 (23%) | 4085 (13%) | |
| 35-49 | 1334706 (21%) | 1326222 (21%) | 8484 (27%) | |
| 50-64 | 1279002 (20%) | 1268832 (20%) | 10170 (32%) | |
| ≥ 65 | 1029846 (16%) | 1022503 (16%) | 7343 (23%) | |
| Race | | | | <0.001 |
| White | 3878526 (60%) | 3858476 (60%) | 20050 (63%) | |
| Black | 834719 (13%) | 829734 (13%) | 4985 (16%) | |
| Hispanic | 785390 (12%) | 781181 (12%) | 4209 (13%) | |
| Asian | 137343 (2%) | 136718 (2.1%) | 625 (2.0%) | |
| Other | 20456 (0.3%) | 20339 (0.3%) | 117 (0.4%) | |
| Unknown | 760066 (12%) | 758308 (12%) | 1758 (5.5%) | |
| Asthma | 445268 (7%) | 439770 (7%) | 5498 (17%) | <0.001 |
| Malignant cancer | 322293 (5%) | 319164 (5%) | 3129 (10%) | <0.001 |
| Diabetes complicated | 315855 (5%) | 312144 (5%) | 3711 (12%) | <0.001 |
| Congestive heart failure | 264487 (4%) | 261211 (4%) | 3276 (10%) | <0.001 |
| Coronary artery disease | 321043 (5%) | 317735 (5%) | 3308 (10%) | <0.001 |
| Hypertension | 1388522 (22%) | 1374529 (22%) | 13993 (44%) | <0.001 |
| Peripheral vascular disease | 143177 (2%) | 141552 (2%) | 1625 (5%) | <0.001 |
| SOT Transplant | 43565 (1%) | 42867 (1%) | 698 (2%) | <0.001 |
| Multiple transplant | 4783 (11%) | 4671 (11%) | 112 (16%) | |
| Kidney transplant | 24129 (55%) | 23813 (56%) | 316 (45%) | |
| Liver transplant | 6998 (16%) | 6933 (16%) | 65 (9%) | |
| Lung transplant | 3068 (7%) | 2946 (10%) | 122 (17%) | |
| Heart transplant | 4587 (11%) | 4504 (11%) | 83 (12%) | |

Table 2: Odd Ratio (95% CI) of the Long COVID outcome from multivariable logistical regression in the acute COVID cohort.

| Variables | Odd Ratio (95% CI) | P value |
|---|---|---|
| Male | Ref. | |
| Female | 1.35 (1.32, 1.38) | <0.001 |
| Age Group | | |
| 19-34 | Ref. | |
| <19 | 0.55 (0.52, 0.58) | <0.001 |
| 35-49 | 2.05 (1.97, 2.13) | <0.001 |
| 50-64 | 2.24 (2.16, 2.33) | <0.001 |
| ≥ 65 | 1.73 (1.66, 1.81) | <0.001 |
| Race Ethnicity | | |
| White | Ref. | |
| Black | 1.02 (1.00, 1.05) | 0.03 |
| Hispanic | 1.18 (1.14, 1.22) | <0.001 |
| Asian | 0.98 (0.90, 1.06) | 0.6 |
| Other | 1.42 (1.17, 1.69) | <0.001 |
| Unknown | 0.59 (0.56, 0.62) | <0.001 |
| Hypertension | 1.36 (1.32, 1.40) | <0.001 |
| Diabetes | 1.09 (1.05, 1.14) | <0.001 |
| Asthma | 1.85 (1.79, 1.91) | <0.001 |
| Malignant cancer | 1.12 (1.07, 1.16) | <0.001 |
| Congestive heart failure | 1.27 (1.22, 1.32) | <0.001 |
| Peripheral vascular disease | 1.02 (0.96, 1.07) | 0.6 |
| Dementia | 1.18 (1.08, 1.29) | <0.001 |
| Transplant Type | | |
| Non-Transplant | Ref. | |
| Multiple | 2.29 (1.88, 2.75) | <0.001 |
| Kidney | 1.33 (1.19, 1.49) | <0.001 |
| Liver | 1.04 (0.81, 1.32) | 0.7 |
| Lung | 4.06 (3.36, 4.85) | <0.001 |
| Heart | 1.72 (1.37, 2.13) | <0.001 |

Table 3: Hazard Ratio of the time-to-event outcome of patient survival from multivariable Cox models in the Long COVID cohort.

| Variables | Hazard Ratio (95% CI) | P value |
|---|---|---|
| Gender | | |
| Male | Ref. | |
| Female | 0.69(0.68, 0.70) | <0.001 |
| Age Group | | |
| 19-34 | Ref. | |
| <19 | 2.74(2.57, 2.92) | <0.001 |
| 35-49 | 3.16(2.97, 3.36) | <0.001 |
| 50-64 | 9.19(8.69, 9.72) | <0.001 |
| ≥ 65 | 27.8(26.3, 29.4) | <0.001 |
| Race Ethnicity | | |
| White | Ref. | |
| Black | 1.13(1.11, 1.16) | <0.001 |
| Hispanic | 1.11(1.08, 1.15) | <0.001 |
| Asian | 1.28(1.21, 1.35) | <0.001 |
| Other | 0.44(0.33, 0.57) | <0.001 |
| Unknown | 1.03(1.01, 1.06) | 0.019 |
| Hypertension | 1.57(1.54, 1.60) | <0.001 |
| Diabetes | 1.34(1.31, 1.37) | <0.001 |
| Malignant | 1.51(1.48, 1.54) | <0.001 |
| Congestive heart failure | 2.69(2.64, 2.74) | <0.001 |
| Peripheral vascular disease | 1.16(1.13, 1.18) | <0.001 |
| Dementia | 1.15(1.11, 1.19) | <0.001 |
| Transplant Type | | |
| Non-Transplant | Ref. | |
| Multiple transplantations | 1.19(1.05, 1.35) | 0.006 |
| Kidney transplantation | 1.87(1.78, 1.97) | <0.001 |
| Liver transplantation | 1.63(1.47, 1.81) | <0.001 |
| Lung transplantation | 2.05(1.80, 2.32) | <0.001 |
| Heart transplantation | 0.91(0.80, 1.03) | 0.14 |

**References**

1. Arab-Zozani, M., Hashemi, F., Safari, H., Yousefi, M., & Ameri, H. (2020, October). *Health-related quality of life and its associated factors in COVID-19 patients*. Osong public health and research perspectives. Retrieved December 11, 2022, from https://www.ncbi.nlm.nih.gov/pmc/articles/PMC7577388/
2. Azzi Y;Parides M;Alani O;Loarte-Campos P;Bartash R;Forest S;Colovai A;Ajaimy M;Liriano-Ward L;Pynadath C;Graham J;Le M;Greenstein S;Rocca J;Kinkhabwala M;Akalin E; (n.d.). *COVID-19 infection in kidney transplant recipients at the epicenter of pandemics*. Kidney international. Retrieved December 11, 2022, from https://pubmed.ncbi.nlm.nih.gov/33069762/
3. Chopra, V., University of Michigan Health System and The Michigan Hospital Medicine Safety Collaborative, Flanders, S. A., O'Malley, M., Malani, A. N., St. Joseph Mercy Health System and The Michigan Hospital Medicine Safety Collaborative, Prescott, H. C., & Tipirneni, R. (2020, November 16). *Sixty-day outcomes among patients hospitalized with COVID-19*. Annals of Internal Medicine. Retrieved December 11, 2022, from https://www.acpjournals.org/doi/10.7326/M20-5661
4. Tabacof L, Tosto-Mancuso J, Wood J, et al. Post-acute COVID-19 syndrome negatively impacts health and wellbeing despite less severe acute infection. bioRxiv. 2020; published online Nov 6. DOI:10.1101/2020.11.04.20226126.
5. Menges D, Ballouz T, Anagnostopoulos A, et al. Estimating the burden of post-COVID-19 syndrome in a population-based cohort study of SARS-CoV-2 infected individuals: Implications for healthcare service planning. 2021; published online March 1 . DOI:10.1101/2021.02.27.21252572.
6. Sheehy LM. Considerations for Postacute Rehabilitation for Survivors of COVID-19. JMIR Public Health Surveill 2020; 6: e19462..
7. Elias M, Pievani D, Randoux C, et al. COVID-19 infection in kidney transplant recipients: disease incidence and clinical outcomes. J Am Soc Nephrol. 2020; 31:2413–2423.
8. Pereira MR, Mohan S, Cohen DJ, et al. COVID-19 in solid organ transplant recipients: initial report from the US epicenter. Am J Transplant. 2020;20:1800–1808.
9. Kates OS, Haydel BM, Florman SS, et al. Coronavirus disease 2019 in solid organ transplant: a multi-center cohort study. Clin Infect Dis. [Epub ahead of print. August 7, 2020]. doi: 10.1093/cid/ciaa1097
10. Miarons, M., Larrosa-García, M., García-García, *et al*. COVID-19 in Solid Organ Transplantation: A Matched Retrospective Cohort Study and Evaluation of Immunosuppression Management. *Transplantation*, *105*(1), 138–150. https://doi.org/10.1097/TP.0000000000003460
11. Vinson, A. J., Dai, R., Agarwal, G., Anzalone, A. J., Lee, S. B., French, E., Olex, A. L., Madhira, V., Mannon, R. B., & National COVID Cohort Collaborative (N3C) Consortium (2022). Sex and organ-specific risk of major adverse renal or cardiac events in solid organ transplant recipients with COVID-19. *American journal of transplantation : official journal of the American Society of Transplantation and the American Society of Transplant Surgeons*, *22*(1), 245–259. https://doi.org/10.1111/ajt.16865
12. Cao, X. *COVID-19: Immunopathology and its implications for therapy*. Nature reviews. Immunology. Retrieved December 11, 2022, from https://pubmed.ncbi.nlm.nih.gov/32273594/
13. Guillen E, Pineiro GJ, Revuelta I, et al. Case report of COVID-19 in a kidney transplant recipient: does immunosuppression alter the clinical presentation? [published online ahead of print 2020]. *Am J Transplant*. 10.1111/ajt.15874
14. Singh, A.K., Gilles, C.L., Singh, R., Singh, A., Chudasama, Y., Coles, B., et al. **Prevalence of comorbidities and their association with mortality in patients with COVID-19: A Systematic Review and Meta-analysis** Diabetes Obes Metab (2020)

15. Caillard S;Chavarot N;Francois H;Matignon M;Greze C;Kamar N;Gatault P;Thaunat O;Legris T;Frimat L;Westeel PF;Goutaudier V;Jdidou M;Snanoudj R;Colosio C;Sicard A;Bertrand D;Mousson C;Bamoulid J;Masset C;Thierry A;Couzi L;Chemouny JM;Duveau A;Moal V;Blancho. (n.d.). *Is covid-19 infection more severe in kidney transplant recipients?* American journal of transplantation : official journal of the American Society of Transplantation and the American Society of Transplant Surgeons. Retrieved December 11, 2022, from https://pubmed.ncbi.nlm.nih.gov/33259686/
16. Caillard, S., Anglicheau, D., Matignon, M., Dürrbach, A., Greze, C., Frimat, L., Thaunat, O., Legris, T., Moal, V., Westeel, P. F., Kamar, N., Gatault, P., Snanoudj, R., Sicard, A., Bertrand, D., Colosio, C., Couzi, L., Chemouny, J. M. M., Masset, C., … Quintrec, M. L. (2020, September 11). *An initial report from the French sot covid registry suggests high mortality due to COVID-19 in recipients of kidney transplants*. Kidney International. Retrieved December 11, 2022, from https://hal.archives-ouvertes.fr/hal-02930165
17. Centers for Disease Control and Prevention. (n.d.). *Underlying medical conditions associated with higher risk for severe COVID-19: Information for Healthcare professionals*. Centers for Disease Control and Prevention. Retrieved December 11, 2022, from https://www.cdc.gov/coronavirus/2019-ncov/hcp/clinical-care/underlyingconditions.html
18. Chan, J. W. M., Ng, C. K., Chan, Y. H., Mok, T. Y. W., Lee, S., Chu, S. Y. Y., Law, W. L., Lee, M. P., & Li, P. C. K. (2003, August). *Short term outcome and risk factors for adverse clinical outcomes in adults with severe acute respiratory syndrome (SARS)*. Thorax. Retrieved December 11, 2022, from https://www.ncbi.nlm.nih.gov/pmc/articles/PMC1746764/
19. Cravedi P, Mothi SS, Azzi Y, et al. COVID-19 and kidney transplantation: results from the TANGO international transplant consortium. Am J Transplant. 2020;20(11):3140-3148.
20. Garg S, Kim L, Whitaker M, O'Halloran A, Cummings C, et al. Hospitalization rates and characteristics of patients hospitalized with laboratory-confirmed coronavirus disease 2019 — COVID-NET, 14 states, March 1–30, 2020. *Centers for Disease Control and Prevention: MMWR.* 2020;69(15):458–464.
21. Guan WJ, Ni ZY, Hu Y, et al; China Medical Treatment Expert Group for COVID-19. Clinical characteristics of coronavirus disease 2019 in China. N Engl J Med. 2020; 382:1708–1720.
22. Guillen E, Pineiro GJ, Revuelta I, et al. Case report of COVID-19 in a kidney transplant recipient: does immunosuppression alter the clinical presentation? [published online ahead of print 2020]. *Am J Transplant*. 10.1111/ajt.15874
23. Hilbrands LB, Duivenvoorden R, Vart P, et al. COV related mortality in kidney transplant and dialysis patients: results of the ERACODA collaboration. Nephrol Dial Transplant. 2020;35(11):1973-1983.
24. Jager KJ, Kramer A, Chesnaye NC, et al.. Results from the ERA-EDTA Registry indicate a high mortality due to COVID-19 in dialysis patients and kidney transplant recipients across Europe. *Kidney Int*. 2020;98:1540–1548.
25. John Hopkins University, https://www.hopkinsmedicine.org/health/conditions-and-diseases/coronavirus/covid-long-haulers-long-term-effects-of-covid19 (2022)
26. Kalyanaraman Marcello R, Dolle J, Grami S, et al.; New York City Health + Hospitals COVID-19 Population Health Data Team. Characteristics and outcomes of COVID-19 patients in New York City's public hospital system. *PLoS One*. 2020;15:e0243027.
27. Kremer D, Pieters TT, Verhaar MC, et al.. A systematic review and meta-analysis of COVID-19 in kidney transplant recipients: lessons to be learned. *Am J Transplant*. [Epub ahead of print. July 1, 2021]. doi: 10.1111/ajt.16742
28. Ledford H. How common is long COVID? Why studies give different answers. Nature. 2022; 606:8523.

29. Michaels MG, La Hoz RM, Danziger-Isakov L, et al. Coronavirus disease 2019: implications of emerging infections for transplantation [published online ahead of print 2020]. *Am J Transplant*. 10.1111/ajt.15832
30. Park, J.E., Jung, A., Park, J.E., MERS **transmission and risk factors: a systematic review** BMC Public Health, 18 (2018), p. 574
31. Pastor-Barriuso R, Perez-Gomez B, Hernan MA, et al. Infection fatality risk for SARS-CoV-2 in community dwelling population of Spain: nationwide seroepidemiological study. BMJ. 2020;371:m4509.
32. RECOVER: Researching COVID to Enhance Recovery. RECOVER: Researching COVID to Enhance Recovery. https://recovercovid.org. Accessed 15 Apr 2022.
33. Singh, A.K., Gilles, C.L., Singh, R., Singh, A., Chudasama, Y., Coles, B., et al. **Prevalence of comorbidities and their association with mortality in patients with COVID-19: A Systematic Review and Meta-analysis** Diabetes Obes Metab (2020)
34. Sheehy LM. Considerations for Postacute Rehabilitation for Survivors of COVID-19. JMIR Public Health Surveill 2020; 6: e19462..
35. Vishnevetsky A, Levy M, **Rethinking high-risk groups in COVID-19** Mult Scler Relat Disord, 42 (2020), p. 102139
36. World Health Organization. WHO coronavirus disease (COVID-19) dashboard, http://covid19.who.int.
37. NIHR. Living with COVID19. Second Review (2021). Available from: https://evidence.nihr.ac.uk/themedreview/living-with-covid19-second-review/ (Accessed March 28, 2021).
38. WHO. A Clinical Case Definition of Post COVID-19 Condition by a Delphi Consensus (2021). Available from: https://www.who.int/publications/i/item/WHO-2019-nCoV-Post_COVID-19_condition-Clinical_case_definition-2021.1 (Accessed October 10, 2021).
39. Zhu L, Xu X, Ma KE, et al. Successful recovery of COVID-19 pneumonia in a renal transplant recipient with long-term immunosuppression [published online ah